\documentclass[aps,10pt,twocolumn,amsmath,amssymb,prl,superscriptaddress]{revtex4-1}

\usepackage{graphicx}
\usepackage{txfonts}
\usepackage[colorlinks,citecolor=blue,linkcolor=blue,urlcolor=blue,plainpages=false,pdfpagelabels]{hyperref}

\newcommand{\LL}{\textrm{LL}}
\newcommand{\CM}{\textrm{CM}}
\newcommand{\rL}{\textrm{L}}
\newcommand{\rR}{\textrm{R}}
\newcommand{\eff}{\!\textrm{eff}}

\begin{document}
\title{Mechanical resonances of mobile impurities in a one-dimensional quantum fluid}

\author{Thomas L. Schmidt}
\email{thomas.schmidt@uni.lu}
\affiliation{Physics and Materials Science Research Unit, University of Luxembourg, L-1511 Luxembourg}

\author{Giacomo Dolcetto}
\affiliation{Physics and Materials Science Research Unit, University of Luxembourg, L-1511 Luxembourg}

\author{Christopher J. Pedder}
\affiliation{Physics and Materials Science Research Unit, University of Luxembourg, L-1511 Luxembourg}

\author{Karyn Le Hur}
\affiliation{Centre de Physique Th\'eorique, \'Ecole Polytechnique, CNRS, Universit\'e Paris-Saclay, F-91128 Palaiseau, France}

\author{Peter P. Orth}
\affiliation{Department of Physics and Astronomy, Iowa State University, Ames, Iowa 50011, USA}
\affiliation{Ames Laboratory, U.S. DOE, Iowa State University, Ames, Iowa 50011, USA}

\date{\today}

\begin{abstract}
We study a one-dimensional interacting quantum liquid hosting a pair of mobile impurities causing backscattering. We determine the effective retarded interaction between the two impurities mediated by the liquid. We show that for strong backscattering this interaction gives rise to resonances and antiresonances in the finite-frequency mobility of the impurity pair. At the antiresonances, the two impurities remain at rest even when driven by a (small) external force. At the resonances, their synchronous motion follows the external drive in phase and reaches maximum amplitude. Using a perturbative renormalization group analysis in quantum tunneling across the impurities, we study the range of validity of our model. We predict that these mechanical antiresonances are observable in experiments on ultracold atom gases confined to one dimension.
\end{abstract}
	
\maketitle

{\it Introduction.} One-dimensional (1D) quantum systems have continued to fascinate condensed-matter physicists over the past decades~\cite{gogolin98,giamarchi03}. They display remarkably greater universality than their higher-dimensional cousins. At the lowest energy scales, and regardless of their microscopic details, or even of whether they are composed of bosons or fermions, all gapless 1D systems can be described as Luttinger liquids (LLs)~\cite{tomonaga50,luttinger63,haldane81}. This hydrodynamic description is extremely simple and characterized by just two parameters: the sound velocity $v_s$ and the Luttinger parameter $K$. The latter determines, e.g., the compressibility, and is related to the interactions between the constituent particles.

In recent years, 1D quantum fluids have been realized on various experimental platforms, ranging from semiconductor nanowires to cold atomic gases~\cite{auslaender05,Steinberg_2007,Barak2010,bloch08,Meinert17}. Many of their predicted properties, such as conductance quantization or the presence of power laws in their response functions (with exponents determined by the interaction strength) have been observed~\cite{bockrath99,ishii03,auslaender05,bloch08,Steinberg_2007}. Cold atoms, in particular, are well-suited for the study of 1D systems as their interaction strength can be tuned via external magnetic fields or by changing properties of the lasers confining the atoms~\cite{bloch08,giorgini08}. Furthermore, it was shown that impurities can be embedded into the 1D quantum fluid, either as atoms of a different species~\cite{Palzer2009,Spethmann-PRL-2012,Catani-PRA-2012,Meinert17} or as charged ions~\cite{Zipkes2010,Sias2014}. Similar effects occur in the propagation of magnetic excitations along spin chains~\cite{Zvonarev-PRL-2007,Zvonarev-PRL-2009,Fukuhara-NatPhys-2013,imambekov12}.

The effects of impurities in 1D quantum fluids are rather peculiar. It is well known that a single static impurity will cause backscattering of fluid particles. At low energies, this backscattering becomes strong for $K< 1$, and the impurity will effectively cut the system into disconnected parts~\cite{kane92,KaneFisher-PRL-1992}. At finite energies, a random scattering potential caused by many static impurities leads to Anderson localization even for arbitrarily weak coupling between the impurities and the liquid~\cite{Giamarchi1988,giamarchi03}. However, these predictions apply only to the cases of static impurities and quenched disorder respectively.

\begin{figure}[b!]
\centering
\includegraphics[width=\linewidth]{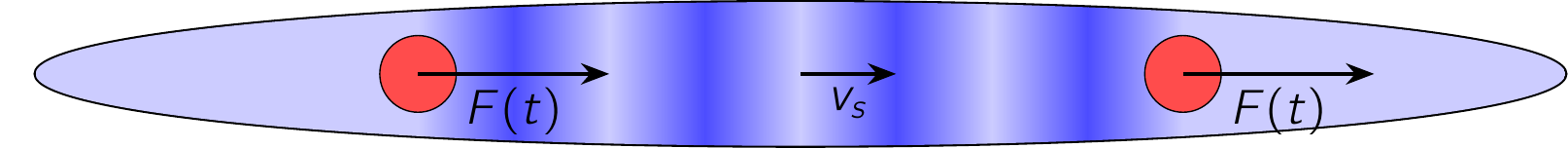}
\caption{Schematic drawing of two impurities immersed in a 1D interacting quantum fluid, which is characterized by the Luttinger parameter $K$ and the sound velocity $v_s$. The two impurities interact with each other via a retarded interaction mediated by the fluid. If exposed to a force $F(t)$, they emit density waves into the liquid as a result of being accelerated. }
\label{fig:1}
\end{figure}

The physical properties of the system change strongly if we allow for backaction of the liquid on the impurities by giving the latter a finite mass $M$~\cite{castroneto96,Lamacraft-PRB-2009,Schecter-AnnPhys-2012,Mathy2012,Massel-NJP-2013,Knap-PRL-2014,Schecter-NJP-2016}. In particular, this gives rise to a new temperature scale $T^* = (m/M) E_F$, where $m$ is essentially the mass of the particles forming the quantum liquid and $E_F$ the Fermi energy \cite{castroneto96}. For $T^* \ll T \ll E_F$, the impurity indeed cuts the system, while at even lower temperatures $T \ll T^*$, the impurity effectively becomes transparent, in striking contrast to the infinite-mass case. The two limits can be distinguished by measuring the impurity's mobility, i.e., its velocity as a response to an external driving force.

Considering more than a single impurity leads to even more striking effects. For static impurities, it is well known that tunneling resonances occur~\cite{kane92}, as well as Friedel oscillations \cite{matveev93,egger95} leading to long-range interactions between two impurities. In the case of mobile impurities, the backaction of the liquid on the impurities necessarily leads to an effective retarded interaction between them. Interactions mediated by a fluid and new dynamical synchronization phenomena have been observed in a cold-atom mixture in Ref.~\cite{Delehaye_2015}. In this paper, we will focus on the mobility of a pair of impurities in the limit of weak tunneling of liquid particles across the impurities. We will show that as a consequence of retarded interactions between the impurities mediated by the liquid, the ac mobility of the pair differs greatly from the mobility of individual impurities. As illustrated in Fig.~\ref{fig:1}, the behavior of the ac mobility is dominated by interference effects between density waves in the liquid and the external driving force. Such behavior is reminiscent of that underlying Fano resonances~\cite{zubairy:qo}; the origin of the two interfering wavefronts, however, is different in our case. The most striking consequence of this is the emergence of resonances and antiresonances in the ac mobility: at certain discrete drive frequencies, the impurity motion is reinforced or completely halted by the density waves in the fluid. We show below that these features of the mobility at finite frequencies can be experimentally observed by measuring the absorption spectrum for electromagnetic radiation.

The resonance phenomena that we discover have the following properties: firstly, it is crucial that the system is one-dimensional because a dispersion of the density waves into other directions would destroy the exact interference of the external force with the force due to the density wave emitted from the second impurity. Secondly, the resonances and antiresonances are retardation phenomena which rest on the finiteness of the sound velocity in the LL and the finite distance between impurities. Thirdly, the leading correction causing a partial lifting of the resonances is brought about by quantum mechanics: both via tunneling of the liquid across the impurity and via quantum fluctuations of their distance $d$. An implication of our result is that one can control the motion of the impurities through the liquid by changing their distance. In particular, one can make the pair immobile at discrete frequencies (up to linear order in the external force) by adjusting $d$. At the same time, the power absorbed when driving oscillatory motion of the pair strongly depends on $d$ as we show below.

The structure of this paper is as follows. Starting from the limit of strong backscattering at the impurities, we will first derive an effective action for the two impurities by integrating out the degrees of freedom of the quantum liquid. This effective action will then contain a retarded interaction between the impurities and we will discuss its effect on the ac mobility. Next, we discuss the possible effects on the real-time dynamics of the impurities, the power absorbed from the external drive, as well as the implications for experiments on cold atom gases. Finally, we will perform a renormalization group (RG) analysis for the tunneling across the impurities to explore the limits of validity of the model.

{\it Effective action.} We begin by presenting a model Hamiltonian to investigate a pair of mobile impurities interacting with an LL. The Hamiltonian $H$ of the system divides into three parts, $H = H_{\LL} + H_{\textrm{imp}} + H_{\textrm{int}}$, which describe the low-energy modes of the liquid, the dynamics of the mobile impurities, and the coupling between the two respectively. $H_{\LL}$ is the usual Luttinger Hamiltonian~\cite{giamarchi03},
\begin{align}
H_{\LL} &= \frac{v_s}{2\pi} \int \, dx \, \left[ K (\partial_x \theta)^2 + \frac{1}{K}(\partial_x \phi)^2  \right],
\end{align}
where $v_s$ is the sound velocity of the fluid modes, and $K$ is the Luttinger parameter, which for repulsive interactions in a fermionic system takes values $0<K<1$. The fields $\phi$ and $\theta$ obey the canonical commutation relation $[\phi(x), \partial_y \theta(y)] = i\pi \delta (x-y)$.

The kinetic energy of the mobile impurities can be written in center-of-mass (CM) and relative ($r$) coordinates as $H_{\textrm{imp}} = P_{\CM}^2/(2M) + P_r^2/(2\mu)$, where $M$ and $\mu$ are the total and reduced mass of the two impurities, respectively. Finally, we allow for a contact interaction between the impurities and the Luttinger modes which is of the form
\begin{align}
    H_{\textrm{int}} &= V \sum_{\eta = \pm} \rho\left(X_{\CM} + \eta X_r/2\right),
    \label{eq:3}
\end{align}
where $\rho = - \partial_x \phi (x)/\pi$ is the density of the LL, and $X_{\CM}$ and $X_{r}$ are the center-of-mass and relative coordinates of the impurity pair.

We transform to a frame comoving with the center of mass of the system using the unitary transformation $U= \exp \left[ i X_{\CM} P_{\LL} \right]$. Here, $P_{\LL} = \int dx (\partial_x \theta) (\rho_0 - \partial_x \phi/\pi)$ is the momentum of the LL, where $\rho_0 = N/L$ denotes the average density. Under this transformation, the Hamiltonian becomes $H^\prime = U^\dag H U$, where
\begin{align}
H^\prime &= \frac{(P_{\CM} - P_{\LL})^2}{2M} +\frac{P_r^2}{2\mu} + H_{\LL} +  V \sum_{\eta = \pm} \rho(\eta X_r/2).
\label{eq:1}
\end{align}
Next, we make two approximations to pass to the Lagrangian formalism. Firstly, we assume that the quantum fluctuations of the relative motion of the impurities are negligible, so that we can approximate the operator by its expectation value $X_r \approx \langle X_r \rangle$, which is justified for large enough impurity mass $M$. We choose to work with external driving forces which affect both impurities identically, so we may further neglect the relative motion and set $X_r \approx d = \text{const}$. For charged impurities, the most straightforward way to implement such a drive is via a time-dependent electric field with wavelength $\lambda \gg d$, as both impurities will experience the same driving force in this case. Secondly, we consider the limit of strong backscattering by the impurities ($V \to \infty$), which leads to complete pinch-off of the LL at the impurities. This is justified because the impurities act as RG-relevant perturbations. We will discuss this assumption later by presenting an RG calculation of tunneling across the impurities.
The Lagrangian then reads
\begin{align}
L(t) &= \frac{1}{2} M \dot{X}_{\CM}^2 + \rho_0 \, \dot{X}_{\CM} \int \, dx \, (\partial_x \theta)  \notag \\
&+  \frac{K}{2 \pi v_s} \int \, dx \, [(\partial_t \theta)^2 - v_s^2 (\partial_x\theta)^2 ],
\end{align}
where we dropped the RG-irrelevant term in $P_{\LL}$. Note that $\theta(x)$ is discontinuous at $x = \pm d/2$ due to the presence of the impurities. We pass to Matsubara space operators (denoted by tildes), and integrate out the LL modes away from the impurity positions at $x=\pm d/2$ to find that in the absence of tunneling across impurities, the action takes the form
\begin{align}
S &= \frac{1}{\beta} \sum_{\omega_n} \Bigg\{ \frac{M \omega_n^2}{2} |\tilde{X}_{\CM}|^2 - \rho_0 \omega_n \tilde{X}_{\CM}^* \sum_{\alpha=\rL,\rR} \sum_{\eta = \pm}\eta \tilde{\theta}_{\alpha,\eta} + \notag \\
&+ \frac{K |\omega_n|}{2\pi} \Bigg[ |\tilde{\theta}_{\rL-}|^2 + |\tilde{\theta}_{\rR+}|^2 \notag \\
&+ \coth \left(\frac{|\omega_n|}{\Omega_d} \right)\left(|\tilde{\theta}_{\rL+}|^2 + |\tilde{\theta}_{\rR-}|^2\right)
 - \frac{\tilde{\theta}_{\rL+}^* \tilde{\theta}_{\rR-} + \rm{c.c.}}{\sinh \left( |\omega_n|/\Omega_d \right)}  \Bigg] \Bigg\}, \label{eq:action}
\end{align}
where $\tilde{X}_{\CM}(\omega_n) = \int dt X_{\CM}(t) e^{ i \omega_n t} $ with bosonic Matsubara frequency $\omega_n = 2 \pi n/\beta$ and inverse temperature $\beta = 1/T$. The frequency $\Omega_d = v_s/d$ is the energy scale of fluctuations of the fluid between the impurities. Moreover, $\tilde{\theta}_{L,\pm} = \tilde{\theta}(-d/2 \pm 0^+)$ and $\tilde{\theta}_{R,\pm} = \tilde{\theta}(d/2 \pm 0^+)$ denote liquid fields near the impurities. Finally, one can integrate out the remaining bosonic modes to find the following effective action for the impurities,
\begin{align}\label{eq:Seff}
    S_{\eff} = \frac{1}{\beta} \underset{\omega_n}{\sum} \left\{ \frac{M \omega_n^2}{2} + \frac{2 \pi\rho^2 |\omega_n|}{K\left[1+ \exp(-|\omega_n|/\Omega_d)\right]} \right\} |\tilde{X}_{\CM}|^2.
\end{align}
The effective action $S_{\eff}$ shows that the two impurities are coupled by a retarded interaction, and thus behave profoundly differently to a single mobile impurity.

{\it Mobility.} The effects of this retarded interaction between the impurities are seen most clearly when considering the mobility of the impurity pair, $\mu(\omega)$, i.e., the response of the CM velocity to an external force acting on both impurities. In a diffusive system, the mobility is the proportionality coefficient between an applied force $\tilde{F}(\omega)$ and the velocity $v(\omega)$ of the particle, $v(\omega) = \mu(\omega) \tilde{F}(\omega)$. The zero-frequency (or dc) mobility reads $\mu_0 \equiv \mu(0) = 1/(M \Omega_M)$, where $\Omega_M = 2\pi \rho^2/(KM )$ \cite{castroneto96}. The inverse of the frequency $\Omega_M$ can be interpreted as the time scale on which a particle with mass $M$ initially at rest approaches its terminal velocity under the application of a constant force. A divergent $\mu_0$ indicates ballistic motion.

\begin{figure}[t]
	\centering
	\includegraphics[width=\linewidth]{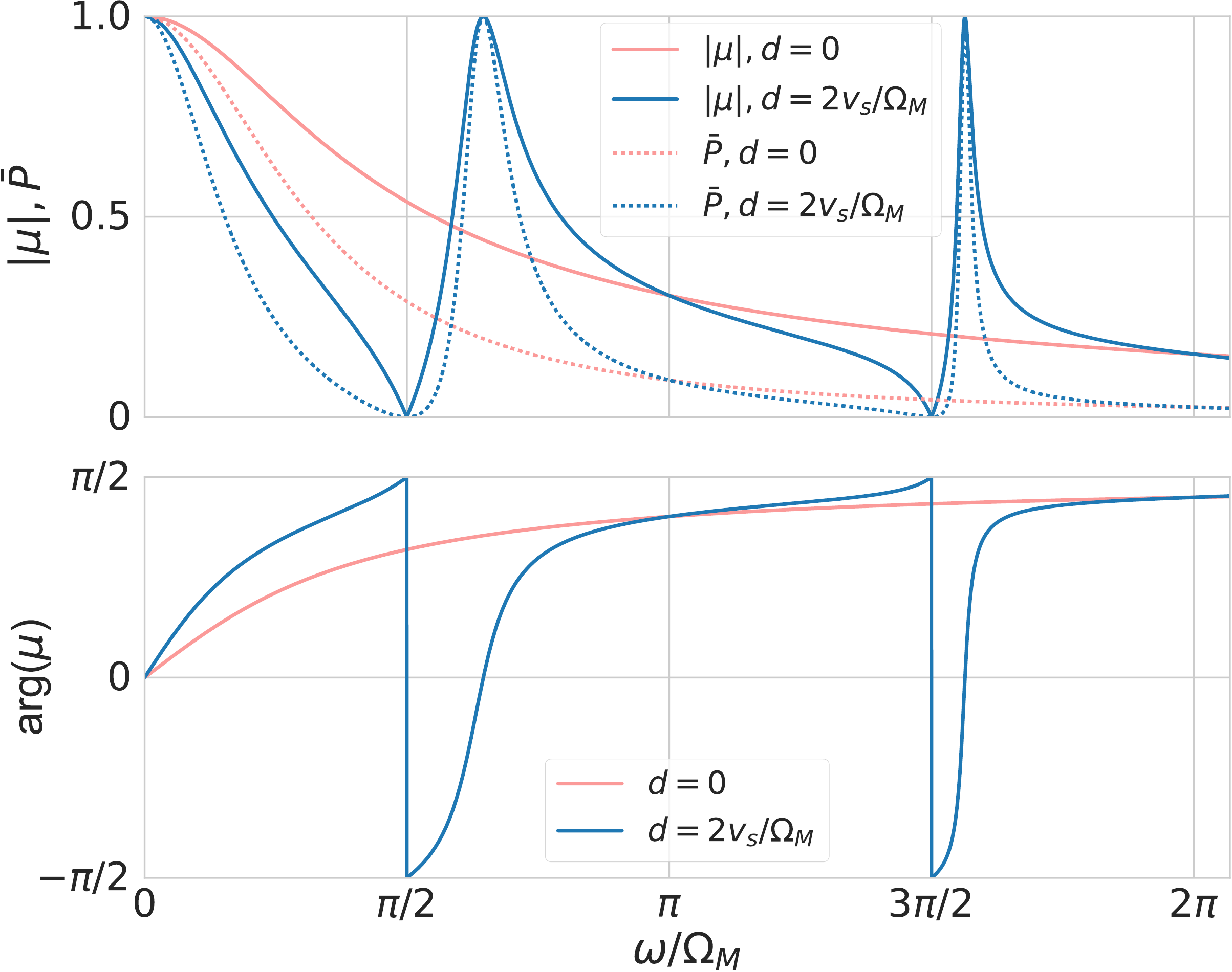}
	\caption{Comparison between the ac mobility $\mu(\omega)$ [Eq.~(\ref{eq:mobility})] and absorbed power $\bar{P}(\omega)$ [Eq.~(\ref{eq:power})] of one impurity ($d = 0$, red curve) and two impurities ($d = 2 v_s/\Omega_M$, blue curve) as a function of frequency. \emph{Top panel:} absolute value of $\mu(\omega)$ and $\bar{P}(\omega)$ normalized with respect to their maxima. \emph{Lower panel:} complex phase of ${\mu}(\omega)$. For a single impurity, the mobility decreases monotonically with $\omega$ while the phase monotonically increases from zero to $\pi/2$. In contrast, for two impurities one finds a series of resonances where $\mu(\omega) =\mu_0$ and $\text{arg}(\mu) = 0$, as well as antiresonances where $\mu(\omega)=0$ and $\text{arg}(\mu) = \pm \pi/2$.}
	\label{fig:2}
\end{figure}

The ac mobility $\mu(\omega)$ can be calculated directly from the effective action (\ref{eq:Seff}), and one finds that the boundary conditions provided by the LL have a strong impact on it. The center of mass mobility as a function of Matsubara frequency is defined as $\mu (\omega_n) = (\omega_n/\beta) \langle | \tilde{X}_{\CM} (\omega_n)^2 | \rangle$. Computing the expectation value, and passing back to real frequencies by making the Wick rotation $\omega_n \rightarrow -i \omega + 0^+$, we find that
\begin{align}\label{eq:mobility}
\frac{\mu(\omega)}{\mu_0} = \left\{1 -i\left[\frac{\omega}{\Omega_M} + \tan \left(\frac{\omega }{2\Omega_d} \right)\right] \right\}^{-1}.
\end{align}
It is clear from this expression that the mobility exhibits \emph{resonances} for $\omega/\Omega_M + \tan (\omega/2\Omega_d) = 0$, where it reaches the dc value: $\mu(\omega) = \mu_0$. Interestingly, it also exhibits \emph{antiresonances} for $\omega/\Omega_M + \tan (\omega/2\Omega_d) \rightarrow \infty$, where $\mu(\omega) = 0$. The full behavior of $\mu(\omega)$ is shown in Fig.~\ref{fig:2}, which shows that the mobility of the impurity pair is very different from that of a single impurity, except at certain discrete frequencies, where they agree.

To understand the form of $\mu(\omega)$, note that the ac drive causes the impurities to emit sound waves into the quantum liquid~\cite{castroneto96}. The antiresonances are reached exactly when the tangent function diverges, i.e., for $\omega = n \pi \Omega_d$ with odd $n$. This corresponds to situations where the driving force at, say, the right impurity and the wavefront arriving with retardation $\Omega_d^{-1} = d/v_s$ from the left impurity interfere destructively. In that case, the forces cancel and the impurities will remain at rest despite the drive, so that $\mu = 0$.

Further increasing the frequency $\omega$, one reaches a point when the external driving force on the impurities and the force due to absorption and emission of liquid density waves conspire in such a way that the impurity motion is always exactly in phase with the external drive, even at finite frequency. This requires that the acceleration of the impurity vanishes exactly at the time when the external force changes direction. In this case, the dc mobility is restored, and the oscillation amplitude as well as the power absorption reach a maximum (see Fig.~\ref{fig:2}). This is a striking contrast to the case of a single impurity, where the motion is always phase shifted with respect to the ac drive.

Finally, for drive frequencies $\omega$ corresponding to even multiples of $\pi\Omega_d$, the two forces interfere constructively, and the mobility for finite $d$ coincides with that of $d=0$. Hence, the two impurities behave identically to a single impurity as far as the mobility is concerned.

{\it Real-time dynamics.} In order to get a complete physical understanding of the system, we investigate the center-of-mass dynamics of the impurities in the time domain. The velocity of the center of mass is given by
\begin{align}
v_{\CM}(t) =  \int  \frac{d\omega}{2\pi} e^{-i \omega t} \mu(\omega) \tilde{F}(\omega) \label{eq:freqeqn}
\end{align}
where $\tilde{F}(\omega) = \int dt e^{i\omega t} F(t)$ is the applied force in frequency domain. From Eq.~(\ref{eq:mobility}), one finds that $\mu(\omega) = \mu^*(-\omega)$ and that the singularities of $\mu(\omega)$ are in the lower complex half-plane, so $v_\CM(t)$ is assured to be real and respect causality.

Our previous analysis of $\mu(\omega)$ in the frequency domain immediately yields the real-time center-of-mass dynamics of the impurity pair for the case of a harmonically-driven system, where $F(t) = F_0 \sin (\omega t)$,
\begin{align}
    v_{\text{CM}}(t) = F_0 |\mu(\omega)| \sin [\omega t - \varphi(\omega)].
\end{align}
where we have split the mobility into its absolute value and phase using $\varphi(\omega) = \arg[\mu(\omega)]$. According to Fig.~\ref{fig:2}, when $\omega$ is on resonance, the velocity $v_\CM(t)$ oscillates in phase with the external force and with the maximum amplitude, $v_\CM(t) = F_0 \mu_0 \sin(\omega t)$. By contrast when $\omega$ hits an antiresonance, we find $v_\CM(t) = 0$ at all times. Near the antiresonances, the relative phase of the center-of-mass velocity with respect to the external drive changes from $\pi/2$ to $-\pi/2$. In the case of a periodic drive, the mobility can most easily be measured by studying the power absorbed by the impurities over a drive period $T = 2\pi/\omega$,
\begin{align}\label{eq:power}
    \bar{P}(\omega) = \frac{1}{T} \int_0^T dt F(t) v_\CM(t) = \frac{1}{2} F_0^2 |\mu(\omega)| \cos[\varphi(\omega)]
\end{align}
Hence, the resonances and antiresonances in Fig.~\ref{fig:2} are directly measurable, for instance, as the power absorption by charged impurities driven by an electric field.

{\it RG analysis.} To obtain these results, we assumed the two impurities to pinch off the liquid at the positions $\pm d/2$. The leading perturbations to this strong-pinning limit are described by adding terms to the action (\ref{eq:action}) corresponding to tunneling across the individual impurities and across the impurity pair. These are given by the respective actions,
\begin{align}
    S_{\rm tun,1}
&=
    -t_1 \sum_{\alpha=L,R} \int d\tau \cos\left[ \theta_{\alpha+}(\tau) - \theta_{\alpha-}(\tau) \right],  \notag \\
    S_{\rm tun,2}
&=
    -t_2 \int d\tau \cos\left[ \theta_{R+}(\tau) - \theta_{L-}(\tau) \right].
\end{align}
We perform an RG analysis for these terms by splitting the bosonic modes into slow modes and fast modes, and integrating out the fast modes \cite{giamarchi03}. The RG flow at the system energy scale $\Lambda$ is determined by the energy $\Omega_d=v_s/d$, corresponding to the distance between the impurities, as well as $\Omega_M=2\pi \rho^2/(K M)$, corresponding to the mass of the impurities.

In the limit of $d \to \infty$, we find the following RG flow for the cases of heavy ($M\rightarrow \infty$) and light ($M\rightarrow 0$) impurities, respectively,
\begin{align}
    \Omega_d \ll \Omega_M \ll \Lambda: \quad &
    \frac{dt_{1,2}}{d\ell} =  \left[ 1 - \frac{1}{K} \right] t_{1,2}, \notag \\
    \Omega_d \ll \Lambda \ll \Omega_M: \quad &
    \frac{dt_{1,2}}{d\ell} =  \left[ 1 - \frac{1}{2K} \right] t_{1,2}.
\end{align}
The tunneling amplitudes $t_1$ and $t_2$ scale identically. The first line coincides with the result for static impurities \cite{kane92} and shows that tunneling is irrelevant for $K < 1$ for heavy impurities. Interestingly, for light impurities, the second line shows that in this regime of large $d$, tunneling is relevant for $1/2 < K < 1$, in contrast to the static case. If the interactions in the LL are not too strong, light impurities thus become transparent even though they are far apart. For sufficiently strong interactions $K < 1/2$, tunneling remains irrelevant. This central, new result of our RG analysis implies that the resonance phenomena discussed in the previous section remain robust to quantum tunneling across the impurities for sufficiently strong interactions in the LL ($K< 1/2$).

We note that the scaling dimension of $1/(2K)$ for two light impurities generalizes to $(N-1)/(NK)$ for $N$ light impurities. This factor arises from the existence of a single gapped mode in the limit $\Omega_M \rightarrow \infty$, which reads $\theta_{L-}-\theta_{L+}+\theta_{R-}-\theta_{R+}$ for $N=2$ and generalizes straightforwardly to general $N$. The other $2N-1$ orthogonal modes remain gapless. As the gapping of this mode occurs only in the approximation of keeping $X_r = d$ constant, we conclude that the (quantitative) scaling dimension $1/(2K)$ is not universal and would change if we allowed relative motion of the impurities. We expect the main qualitative conclusion, however, that pairs of light impurities with large $d$ become transparent at smaller $K$ than single impurities, to remain valid.

To connect to the existing results for a single mobile impurity, we need to consider the case $d \to 0$. In that case, the bosonic modes $\theta_{L+}$ and $\theta_{R-}$ which live between the impurities are pushed to high energies. As a consequence, the RG calculation predicts that the tunneling amplitude $t_1$ is exponentially suppressed at energies $\Lambda \ll \Omega_d$. In contrast, the tunneling amplitude $t_2$ remains finite and, in the limits of heavy and light impurities, respectively, scales as,
\begin{align}
    \Omega_M \ll \Lambda \ll \Omega_d: \quad&
    \frac{dt_2}{d\ell} =  \left[ 1 - \frac{1}{K} \right] t_2, \notag \\
    \Lambda \ll \Omega_M \ll \Omega_d: \quad&
    \frac{dt_2}{d\ell} =  t_2.
\end{align}
Hence, we recover the typical scaling in the limit of heavy impurities \cite{kane92}. Moreover, in the limit of light impurities and $d\to 0$, tunneling becomes relevant irrespective of interaction strength. This coincides with the result found in Ref.~\cite{castroneto96} for a single mobile impurity.

{\it Conclusions.} In this paper, we have shown that a pair of mobile impurities embedded in a Luttinger liquid can display rich physical behavior. In particular, we showed that interference effects between an external driving force and forces caused by density waves that are generated in the liquid as a response, can enhance or attenuate the ac mobility of the center-of-mass motion of the two impurities. In the extreme case, antiresonances arising from the destructive interference between external and LL mediated forces can render the impurity pair completely immobile at certain discrete frequencies. We expect that our predictions can most conveniently be measured on ultracold atom gases in one-dimensional traps. In this case, with typical sound velocities in the range of $\text{mm/s}$ and distances between impurities on the order of $\mu\text{m}$, the resonance phenomena should be visible in the electromagnetic absorption at $\text{kHz}$ frequencies.

\begin{acknowledgments}
T.L.S., G.D., and C.J.P.~acknowledge support by the Fonds National de la Recherche Luxembourg under grant ATTRACT 7556175. P.P.O.~acknowledges support from Iowa State University Startup Funds. K.L.H.~acknowledges support by the Deutsche Forschungsgemeinschaft via grant DFG FOR2414 and by the LABEX PALM Grant ANR-10-LABX-0039, as well as discussions at the CIFAR meeting in Canada on quantum materials.
\end{acknowledgments}

\bibliography{refs}
	
\end{document}